# Object-oriented approach to Rapid Custom Instruction design

Emna Kallel, Yassine Aoudni and Mohamed Abid
CES Laboratory, Univ. Sfax, ENIS School
BP 1173, Sfax 3038, Tunisia

*Abstract*—Due to continuous evolution of Systems-on-Chip (SoC), the complexity of their design and development has augmented exponentially. To deal with the ever-growing complexity of such embedded systems, we introduce, in this paper, an object-oriented approach to rapid SoC design using auto-generation of hardware custom instructions to simplify and accelerate the SoC design process. In our approach, a Data Flow Graph (DFG) is adopted as a representation of the arithmetic operation to convert it to a custom instruction. Then VHDL code will be automatically generated. The input C code is automatically updated for calling the new hardware components. To prove the effectiveness of the proposed approach, a Java source code framework named Automatic Custom Architecture generator (ACAgen) is developed. Experimental results on 3D sample application validate our approach and demonstrate how the proposed framework facilitates and accelerates the SoC design process at low costs.

I. INTRODUCTION

General-purpose processors that are utilized as utilized as cores are often incapable of achieving the challenging cost, performance, and power demands of high-performance audio, video, and networking applications. Systems-on-chip often use hardware accelerators or coprocessors to provide efficient implementations of these applications [1]. Indeed, adding application-specific custom hardware in processor core is a method for providing enhanced performance in SoC [2].

However, with the diversity of these techniques, the SoC designer's task will be more complicated in the presence of coprocessor generation and custom-instruction integration problems. In fact, while these specific systems provide good performances, they require long design cycles. The size and complexity involved in their design are continuously outpacing the designer productivity. An important challenge is to find new methodologies that efficiently address the issues about large and complex SoC.

The automatic code generation (ACG) notion has appeared with some techniques to simplify the task of writing a code [3]. It is considered as an efficient solution that allows fast and simple hardware/software co-design. For example, code transformation can represent a high abstraction level, where a part of the code is translated from a source language into a target language. Significant research has been done in automating the hardware generation. For example, approaches based on Model-Driven Engineering (MDE) [4] have been proposed as a solution for complex embedded systems design [5], [6]. Moreover, hardware code generators based on parser generator exist [3], [7] to simplify the SoCs designers tasks. Using these tools, designers can rapidly implement its SoCs. These code generators, however, still require knowledge of hardware design to generate complete and low-cost SoCs .

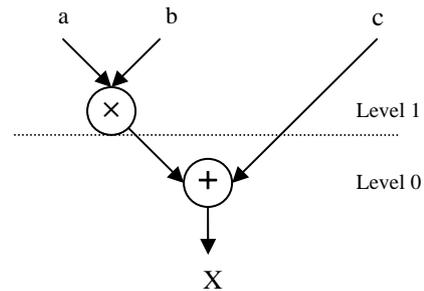

Figure 1. Simple data flow graph (DFG) example

Automatic Custom Architecture generator framework (ACAgen) has been developed to reduce the high proficiency need in hardware development by providing object-oriented methods for Hardware and Software full code generation from a simple user specification. A module of ACAgen framework is developed to automatically integrate custom instructions (CIs) within NIOSII [8] processor core from a DFG. CIs are implemented for the identified parts of the program where it is possible. The identified arithmetic operations are provided in a Data Flow Graph (DFG) such as the one shown in fig.1 before the automatic VHDL code generation. The framework uses a Library of Parameterized Modules (LPM) [9] which offers the convenience of performing mathematical operations on FPGA through parameter function such as adders and dividers. In order to automatically generate complete VHDL code for CIs, two java packages, VJP (VHDL JAVA Package) and LJP (LPM JAVA Package) are implemented in the ACAgen framework.

One original aspect of this project is, on the one hand, the use of CIs to off-load the computationally demanding portions of the application for providing performance improvement and, on the other hand, the development of object-oriented methods to raise the design abstraction level and to reduce the CIs design complexity.

The next section describes the object-oriented approach to CI code generation. It also introduces the developed ACAgen module. Section 3 presents the FPGA based synthesis results of a 3D synthesis application and discusses the experimental results showing the performance of the ACAgen framework. Finally, we end up with a conclusion.

## II. OBJECT-ORIENTED APPROACH TO AUTOMATIC VHDL CODE GENERATION

### A. Overview of ACAgen module for VHDL code generation from DFG

The presented module of ACAgen java Netbeans-based framework takes as entry a DFG and creates synthesizable VHDL code to map it to a custom hardware. The VHDL code generated is adapted to Altera's products and is in a format to be integrated to a Nios II processor as a CI. Nios II is a soft core that offers the possibility to integrate 5 CIs using registers: (dataa[0..31], datab[0..31]) as inputs and (result[0..31]) as output. The CI is implemented using Altera's LPM functions. LPM functions are available in the development tools provided by most FPGA manufacturers to speed up hardware development and include many common components used in VHDL design for arithmetic operations and many other elements. The design flow of VHDL automatic code generation for CI is described in figure 2.
In the proposed design flow, the following main JAVA classes are implemented.

- DFGParser.java is developed to analyze and classify the data of the DFG. It contains functions to separate the DFG data, analyze and list the data and other functions to get the data.
- ComponentMaker.java is designed to generate the hardware LPM components associated to the DFG using the implemented LPM JAVA package.
- VHDLGenerator.java is designed to generate the VHDL code for different parts of the VHDL code using the implemented VHDL JAVA package.

### B. Analysis and Separation Data Flow Graph Information

The first step of the VHDL generation section is to recover and extract the DFG information. Figure 1 shows a DFG for the following simple operation:

$$X = (a * b) + c$$

This information however comes encoded in a string form and must be analyzed and separated. The DFGParser.java class contains some functions that are developed to effectuate several principal tasks such as the analysis, separation, and scheduling of the DFG data. DFGParser.java takes the DFG string, extracts and organizes the information, and then schedules the operations so that the DFG can be created in hardware. After analysis, the DFG data is separated into two ArrayLists DFG operands and operation that are organized by order of nodes priorities. Indeed, the node with a higher level gets the higher priority. For example, the analysis result of the previous DFG example (figure 1) is:

- Sequence of operations: *, +
- Sequence of operands: a, b, c

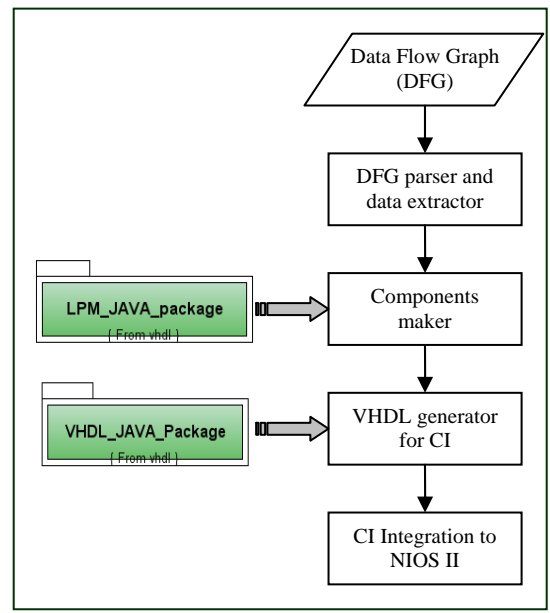

Figure 2. VHDL generation from DFG design flow

### C. LPM components creation

ComponentMaker.java class analyzes the ArrayLists DFG data to find the global inputs to the CI, the interior I/O registers, and the global output of the CI. ComponentMaker.java also figures out what LPM components need to be defined as and how many input stages are needed. The defining of each LPM component should only be done once. There are also some components that are used by multiple operations but have different generics. The LPM_ADD_SUB component is used for both the ADD and SUB instruction. The LPM_DIVIDE module is also used for DIVS, DIVU, MODS, and MODU. The LPM_MULT component is used for the MUL instruction. The operation for these components is specified when the generic and port map is specified. ComponentMaker.java creates an Arraylist of the components that need to be defined.
The mega-function LPM_ADD_SUB does not accept inputs with different sizes, that is why, we have worked on the realization of a VHDL entity that is able to concatenate the bits to adapt them to the desired input.
The concatenation operator combines one-dimensional arrays (or scalars that compose such arrays) to form wider arrays. This operator reverses the effects of slicing arrays. The modulus and remainder operators only differ given negative operands. The remainder has the sign of its dividend;

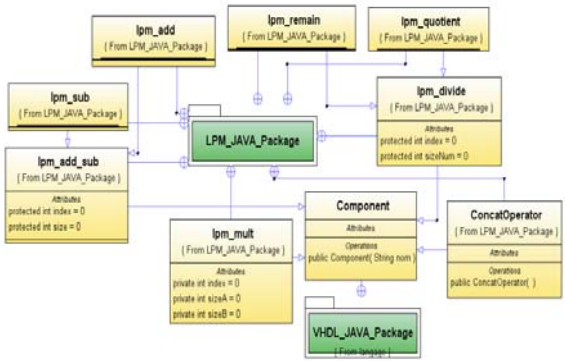

Figure 3. LJP class diagram

the modulus has the sign of its divisor. In both cases, the absolute value of the remainder or modulus is less than the absolute value of the divisor.

In some cases, the concatenation operator and other LMP components can be used several times. It is why, a generic JAVA Package named LPM JAVA package (LJP) is developed to generate the LMP components by just redefining input parameters size. Figure 3 shows the main classes of LPM JAVA package.

### D. VHDL generation

The VHDL code is created by using the data and analysis information collected from the functions of DFGParser.java and ComponentMaker.java. VHDLGenerator.java is developed for instantiating VHDL objects from the VHDL JAVA package (LJP) that brings together all VHDL language aspects to facilitate the VHDL code generation. Figure 4 shows a part of VJP UML class diagram. VJP takes care of the top-level VHDL components generation like the header, the entity, architecture body…

Entity and Architecture are the two main basic programming structures in VHDL. The header of the VHDL contains the libraries to be included and the Entity definition of the CI, which is the standard used for a CI that is added NIOS II processor instruction set. Inputs and outputs of the CI are defined in Entity. In Architecture, behavioral or structural models can be used either to describe the system. Indeed, Architecture is composed of components, signals, port map, connections, processes, etc.

The LPM components that are used and need to be defined is determined earlier using ComponentMaker.java. Moreover, as explained before in section 2.C ComponentMaker.java defines also which cases the concatenation operator needs to be applied. The LPM components classes of LPM JAVA package can now be instantiated. One component is instantiated for each operation and the generics are created. The signals used in the CI are then defined. The port map is also accomplished in this phase and the instantiated LPM ports are mapped to the corresponding I/O port. The function is called for each operation and has inputs for op number, inputA, inputB, and output of the LPM component. The actual operation used is defined in the generic mapping.

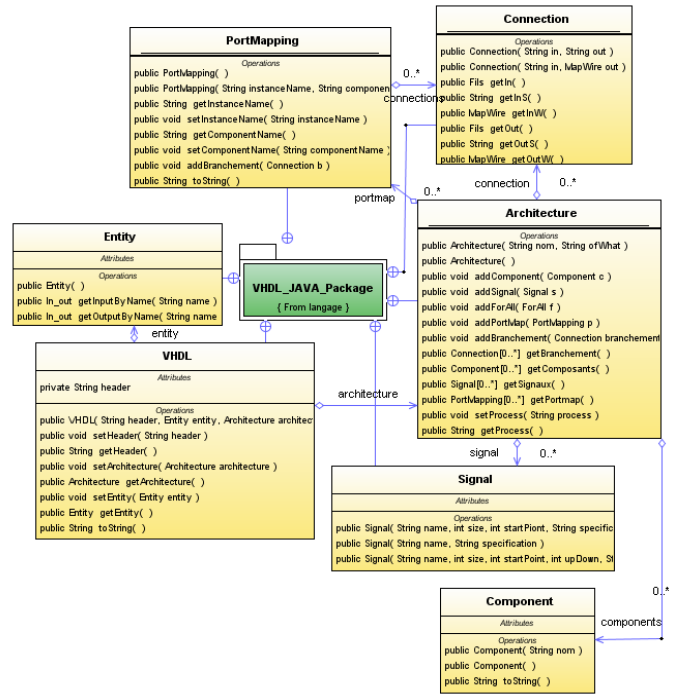

Figure 4. VJP class diagram

The generated VHDL code for the CI follows the specifications described in [10] for the NIOS II processor. The CI input ports are named dataa and datab and the output port is named result. The generated CIs are multicycle thus they use the clk, clk_en, reset, and start inputs along with the done output. This naming convention is done automatically by the ACAgen VHDL generation framework and requires no input from the user. Information on each stage of the VHDL code generation is given in the output console in Netbeans.

### E. CI integration to Nios II

Once the FPGA has been programmed, a C program is automatically updated to call the CI opcode using a developed C parser [11]. The NIOS II IDE is used for developing the software to run on the new hardware system.

### III. EXPERIMENTAL RESULTS

To prove the efficiency of the proposed object-oriented approach, we tested some programs using ACAgen framework. These programs are identified in a C-based 3D graphic pipeline application [2]. NIOSII processor core and StratixII FPGA device are used to prototype the customized reconfigurable SoC. Table I summarizes the results of our experiments. It compares the implementation design time using the framework with results obtained from a conventional manual implementation method done by the same designer without using any frameworks. The results indicate that our approach can quickly (in several seconds to a minute) generate CIs for realistic programs. The selected CIs can achieve an average speedup of 6.78X and peak speedup of 9,6X over the manual solution. Moreover, the power dissipation consumed by the global 3D application is less than 300 mw (table 2).

FAIBLE TENSION FAIBLE CONSOMMATION. IEEE. 2012. (FTFC 2012)

TABLE I. PERFORMANCE IMROVEMENT USING ACAGEN

| Program | # CIs | Design time (H) | | Performance improvement |
|---|---|---|---|---|
| | | *Manual* | *Automatic* | |
| Scalaire | 1 | 165 | 24 | 6.87X |
| Vectoriel | 2 | 230 | 25 | 9.58X |
| Mult_matrice | 1 | 168 | 24 | 7X |
| Projection | 2 | 200 | 25 | 8X |
| Transformation | 3 | 240 | 25 | 9.6X |
| Znormal | 2 | 206 | 25 | 8,24X |

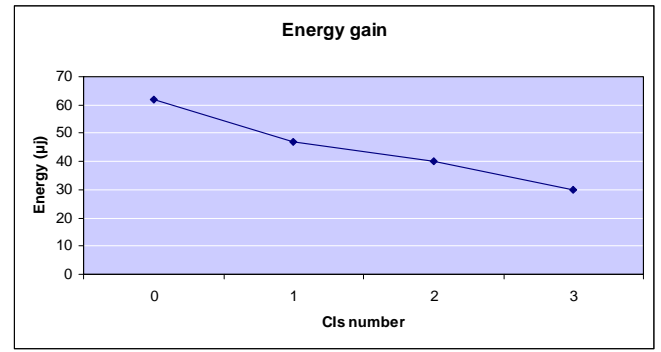

Figure 5. Energy Gain

TABLE II. SYNTHESIS RESULTS

| # CIs | Execution time T(ms) | Resource usage (%) | Power consumption P (mw) | Energy consumption E=PxT (µj) |
|---|---|---|---|---|
| 0 | 31 | 21 | 200 | 62 |
| 1 | 21 | 27 | 223 | 47 |
| 2 | 15 | 32 | 266 | 40 |
| 3 | 10 | 40 | 298 | 30 |

Also, table 2 presents synthesis and execution results on various custom instruction designs. We notice a compromise between area and execution time. The results prove the performance of the proposed design. Indeed, when increasing the number of CIs the speedup increases (3 times higher), the FPGA area is multiplied by a factor of 2 and the power consumption by a factor of 1.5 which are acceptable rates. Moreover, as shown in figure 5, a significant gain in terms of energy consumption is attained when increasing the number of CIs (2 X). This is explained by the increased speedup offered by our system. Indeed, a low power device operating for a long time can use more energy than a high power device operating for a short time. The results show that the FPGA based implementation is inexpensive and can be easily reconfigured.

IV. CONCLUSIONS

In this paper, we proposed an approach to custom instructions automatic generation from a DFG. Our automatic code generation approach is based on objects creation by developing two java packages that bring together all VHDL language aspects and its needed components.

For experimentation, we proposed to use NIOSII processor core and StratixII FPGA device to prototype the customized reconfigurable SoC. 3D sample application was chosen to validate the object-oriented approach. It proved that the developed ACAgen framework can facilitate and accelerate 8x average the custom instruction design and allow reducing implementation costs.